\documentclass[letterpaper]{article}
\usepackage{natbib,alifeconf,algorithm,graphicx,subfig}

\title{How to Measure Group Selection in Real-world Populations}
\author{Simon T. Powers, Christopher Heys \and Richard A. Watson \\
\mbox{}\\
Natural Systems Group, ECS, University of Southampton \\
stp2@ecs.soton.ac.uk}

\begin{document}
\maketitle

\begin{abstract}
Multilevel selection and the evolution of cooperation are fundamental to the formation of higher-level organisation and the evolution of biocomplexity, but such notions are controversial and poorly understood in natural populations. The theoretic principles of group selection are well developed in idealised models where a population is neatly divided into multiple semi-isolated sub-populations. But since such models can be explained by individual selection given the localised frequency-dependent effects involved, some argue that the group selection concepts offered are, even in the idealised case, redundant and that in natural conditions where groups are not well-defined that a group selection framework is entirely inapplicable. This does not necessarily mean, however, that a natural population is not subject to some interesting localised frequency-dependent effects -- but how could we formally quantify this under realistic conditions? Here we focus on the presence of a Simpson's Paradox where, although the local proportion of cooperators decreases at all locations, the global proportion of cooperators increases. We illustrate this principle in a simple individual-based model of bacterial biofilm growth and discuss various complicating factors in moving from theory to practice of measuring group selection.   
\end{abstract}

\section{Group selection in theory and practice}
Some argue that the theoretic principles of group selection are well developed and crucial for understanding evolution in natural populations \citep{Wilson:2007:a,Okasha:2006:a}. Indeed, many artificial life models seeking to explain the evolution of cooperation make either explicit or implicit reference to group-level selection (e.g., \citealt{Scogings:2008:a,Goldsby:2009:a,Wu:2009:a}). The group selection position, however, suffers from at least two serious problems. The first is whether the phenomena involved, though undisputed, formally require group selection concepts. The second is whether the idealised conditions they assume are applicable in natural populations. We briefly overview the standard model of multilevel selection and discuss these limitations. Our aim is to devise a practical theoretical approach to assess whether something interesting is happening in a natural population with respect to the scale of selection. As a practical exemplar, we have in mind the possibility of group selection occurring within natural bacterial biofilms. Biofilms are formed when bacteria attach to a surface and develop into dense aggregations, and they are in fact the most common mode of bacterial growth (compared to well-mixed planktonic populations). Bacteria living in biofilms are known to engage in many cooperative interactions, including the sharing of various `public goods' such as extra-cellular enzymes. Biofilms also exhibit collective properties, such as anti-biotic resistance, that are significantly different from those of free-living bacteria \citep{Ghannoum:2004:a}. Accordingly, they have potential to serve as an ideal model empirical system for studying the transition to multicellularity \citep{Penn:2008:a}. However to do so, we need to be able to connect idealised models of multilevel selection (for example, where groups are discrete and non-overlapping) with real-world biological systems (where the ``groups'' may simply be local neighbourhoods with no discrete boundary). In this paper, we discuss the theoretic and practical issues involved in studying multilevel selection in biofilms and other natural populations. We illustrate our discussion with a simple individual-based model of bacterial growth, in which growth rate depends upon the local concentration of a `public good' that is costly to produce. As such, this system might be expected to fit standard theory on the evolution of cooperation. However in our individual-based model, as in many real-world cases, the groups are not discrete and so it is not immediately obvious how, if at all, a multilevel selection framework can be useful. How, for example, can we measure the relative strengths of within- and between-group selection if the groups do not have discrete boundaries? 

Despite this practical difficulty, theoretical and philosophical work suggests that multiple scales of selection should still be present in such systems \citep{Wilson:1980:a,Sober:1998:a,Nowak:1992:a}. Here, we illustrate the use of Simpson's Paradox \citep{Simpson:1951:a, Sober:1998:a} as a quantifiable indicator of a group-level selection effect. Crucially, we illustrate that this need not rely on a priori knowledge of the exact group structure, or even on the presence of discrete group boundaries. A Simpson's paradox occurs when, although the proportion of cooperators \emph{decreases} in every locality, the global proportion of cooperators nevertheless \emph{increases}. This can be measured in situ and does not require comparison with a well-mixed population, nor that we know the exact evolutionary game (fitness function) that individuals are engaged in. Then, by measuring the magnitude of the discrepancy between local and global proportions of cooperators over a range of local scales, we can identify the effective selective scale in a natural population. We also illustrate several further complicating factors that arise in moving from idealised theoretic models to more realistic biological scenarios.

\section{The idealised model of multilevel selection and its limitations}
The idealised model of multilevel selection involves a population of individuals that is divided into discrete (equal-sized) sub-populations or demes \citep{Wilson:1980:a,Sober:1998:a}, Fig. 1. 

\begin{figure}[h]
\centering
\includegraphics[scale=1]{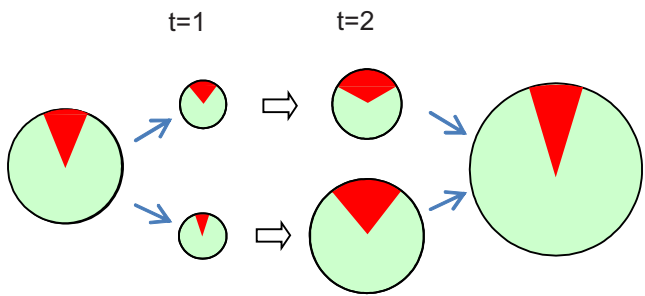}
\caption{Growth of cooperators (green) \& selfish individuals (red) living in groups. Individuals in each group (only two are depicted) are drawn randomly from a global population (left) such that the proportions of types (cooperators and defectors) varies slightly between groups. Groups with more cooperators grow more than groups with fewer cooperators and therefore contribute more individuals (specifically cooperators) to the global cell-count. Hence, the global proportion of cooperators increases (right).}
\label{figPies}
\end{figure}  

Note this model assumes that localised fitness interactions are contained within neatly circumscribed groups. To sustain cooperation at high levels the population must be subject to multiple episodes of `aggregation and dispersal', alternating between phases with a single `migrant pool' (the global population or a representative sample thereof), and phases with multiple localised interaction groups. Without a group mixing stage, selfish behaviour would eventually go to fixation within each group founded by one or more selfish individuals (assuming Prisoner's Dilemma cooperative interactions; \citet{Powers:2008:a,Powers:2010:a}). 

\subsubsection{Is this really group selection? }
It has been widely argued that this classic model shows nothing more than individual selection given localised frequency-dependent effects \citep{Smith:1976:a, Nunney:1985:a, Sterelny:1996:a, Grafen:1984:a}, and hence does not involve group selection at all. That is, rather than saying \emph{groups} with more cooperative individuals are fitter than \emph{groups} with fewer cooperative individuals, we could equally say that \emph{individuals} in groups with more cooperators are fitter than \emph{individuals} in groups with fewer cooperators. In fact, our position is that if we could not explain the outcome of such models in terms of (context dependent) individual selection the result would be `mystical' - that is, we would not have an evolutionary explanation at all. The behaviour of such models is fully explainable, as it must be, in terms of modified selective pressures on individuals given the group-living assumed. Nonetheless, it is at least interesting to note that the increase in levels of cooperation are consistent with the differential productivity of groups, i.e., more cooperative groups are fitter in terms of the genetic contribution they make to future generations, as well as consistent with the differential productivity of individuals, i.e. individuals in more cooperative groups are fitter in terms of the genetic contribution they make to future generations \citep{Dugatkin:1994:a, Kerr:2002:a}. Indeed, this has to be the case because in this (very common) kind of multilevel selection model, group fitness is by definition the mean individual fitness of the group members \citep{Damuth:1988:a, Okasha:2006:a}. However, even this pluralist position seems to be on shaky ground when the groups are not neatly defined. For example, how can we empirically measure group phenotypes (e.g., level of cooperation within the group) if we cannot identify discrete groups? In this case, a group based account will lose accuracy whereas the individual selection perspective remains undeniably precise \citep{Godfrey-Smith:2006:a}.

\subsubsection{Is the standard model relevant to natural populations? }
The standard model describes neatly partitioned sub-populations where the benefits of cooperative acts are distributed equally to members within each group, but not with members of other groups \citep{Wilson:1980:a,Godfrey-Smith:2006:a}. Such idealised conditions are likely to be rare in natural populations. Of course, the effect does not immediately vanish when groups are less neat.  But in such cases, localised frequency-dependent selection seems a perfectly adequate explanation \citep{Smith:1976:a}, and there seems to be little value in arguing for a `group selection' account. Moreover, even if we wanted to retain a group selection framework, it is not clear how we could measure and quantify the differential productivity of groups in realistic scenarios where groups are somewhat ill defined \citep{Godfrey-Smith:2006:a}.

These considerations should not lead one to conclude, however, that there is nothing of consequence presented in the idealised models \citep{Okasha:2006:a} nor that nothing interesting can happen in natural populations. But it is a bit tricky to say what it is exactly, and more tricky to know how to measure it in a natural population. Certainly, if we were to assess the level of cooperation in a natural population, and then (assuming this were practically possible) assess it again in an artificially well-mixed version of the same experiment, we might see a difference in the two levels. This would at least tell us that localised frequency-dependent effects were significant in this system. But frankly, it does not sound all that interesting -- it is rather obvious that selective pressures will be different in well-mixed populations if locally dispersed resources or public goods are involved. Simply examining the global frequency of cooperation tells us nothing about the mechanism behind its evolution, e.g., is cooperation a simple mutualism or is it individually-costly? 

Moreover, although a comparison of well-mixed versus spatial or viscous populations is possible in synthetic simulations, the practicalities of say, mechanically mixing a biofilm or adding surfactants to break-up the extra-cellular matrix that holds cells together would not merely alter spatial relationships, but potentially affect many important environmental factors that could confound the result. We are left, therefore, with a significant gap between the theoretic idealisations of group selection and methodology that would be useful in practical situations \citep{West:2008:a}. 

An alternative is to look for a Simpson's paradox in situ. A Simpson's paradox clearly emphasises the crucial mechanics of multilevel selection \citep{Sober:1998:a}, see below, and it can be measured in situ so that it does not require disruption of the natural population structure. 

\section{Group selection and Simpson's paradox}

Simpson's paradox is a statistical phenomenon that arises when correlations or trends within sub-groups of a data set fail to represent the overall correlation when all the data is assessed together \citep{Simpson:1951:a,Sober:1998:a}. Table 1 shows a very simple hypothetical example based on a group selection scenario. It shows the numbers of cooperators and selfish individuals in two groups, A and B, at two time points, $t=1$ and $t=2$. Note that both groups show a \emph{decrease} in the proportion of cooperators in this time interval, yet overall, from the same data, there is nonetheless an \emph{increase} in the total proportion of cooperators.

\begin{table*}[!htb]
	\centering
	
		\begin{tabular}{|l|lll|lll|}
		\hline
		&&$t=1$&&&$t=2$& \\ \hline
		&Coop&Selfish&\%Coop&Coop&Selfish&\%Coop \\ \hline
		A&2&4&\textbf{33}\%&4&9&31\% \\ 
		B&4&2&\textbf{66}\%&16&10&62\% \\ 
		Total&6&6&50\%&20&19&\textbf{51}\% \\ \hline
		 	
		\end{tabular}
	\caption{Numbers of cooperative and selfish individuals in two hypothetical groups, illustrating Simpson's paradox. Bold highlighting indicates the time point where the proportion of cooperators is highest. Note that within both group A and group B the proportion of cooperators decreases over this period, but overall, the proportion of cooperators increases.}
	\label{tabMinModParams}
\end{table*}

It may be useful to clarify that at a given point in time, the average within-group proportion of cooperators can be different from the global proportion of cooperators. This is simply because the average within-group proportion weights all groups equally, whereas the global proportion is implicitly the same summation but with each group contribution `weighted' in proportion to its size. In the example, at $t=1$ the groups are equal sized and the average within-group proportion and the global proportion are therefore the same. But in the second time point, the groups are different sizes and the average within-group proportion ($(31\%+62\%)/2=46.5\%$) is not equal to the global proportion ($51\%$).

In this example then, the growth trend paradox (i.e., cooperation decreases within groups but increases globally) is caused by the fact that one group grows much more than the other. Specifically, the B group, with twice the initial proportion of cooperators, is assumed to grow at about twice the rate as the A group in this example. So, although selfish individuals always grow faster than the cooperators in any given environment, some cooperators grow faster than some selfish individuals (specifically, when cooperators are in an environment of many other cooperators). Accordingly, because highly cooperative groups grow more, cooperators can increase in total proportion even though they decrease in proportion within each group.

\section{Using Simpson's paradox to indicate group selection}

Simpson's paradox as a basis for group selection is well understood. However, it is generally not used as a direct indicator of group selection. Instead, the norm is simply to assess the global level of cooperation and see if it increases. But in practical experiments this is insufficient to conclude that group selection is responsible for such an increase. When the exact form of the evolutionary game that individuals are engaged in is unknown, due to numerous modes of interaction and multiple `public goods' for example, or competition for multiple resources, it can be difficult to genuinely ascertain whether the `cooperator' is really cooperating and whether the `selfish' type is really selfish. That is, should we be surprised that the global level of cooperation increases, or is it a simple case of mutualism? The obvious control is to compare with a well-mixed population or to increase the diffusion rate in a spatial model, but aside from the practical difficulties of this in natural populations (even bacterial ones), this cannot maintain the `all other things being equal' condition necessary to determine that only the localisation of interactions is producing the difference in results. Instead, by looking for a divergence between the average within-group and global proportions of types, we can both verify that the types are behaving as expected (that in any given environment the selfish individuals have the advantage) and identify a group selection effect if there is one. Thus Simpson's Paradox provides an in situ measurement of group selection in the sense that we do not need to disrupt groups to provide a control, and can therefore assess the effect that groups are having merely by observing how the frequencies of types change in the natural population.

To measure Simpson's Paradox in scenarios that have poorly defined groups requires an additional small step. For this we propose the following practical methodology for a spatially distributed population. Rather than attempt to define boundaries around one group and distinguish it from another, we can simply divide the physical space into equal-sized local regions and measure both the average local proportion of cooperators within all regions, and the global proportion of cooperators. If the selfish individuals are indeed selfish individuals then the average local proportion of cooperators must be always declining. But if, at the same time, the global proportion of cooperators is increasing then there is significant group selection activity. 

Note that if every region exhibited approximately the same amount of total cell growth, then a paradox could not occur; but if some local regions are growing much faster than others (because local frequency-dependent fitness effects are sufficiently strong) a Simpson's Paradox may be observed. In principle, it does not matter whether the space is divided into contiguous tiles (as we employ below), or whether regions are selected at random with random centres. But it does matter that regions are not selected in any manner that is biased by cell density, for that would amount to taking a weighted average. Taking a weighted average would necessarily make the local average the same as the global, and so would result in the local group dynamics disappearing from the analysis. This is the ``averaging fallacy'' described by \citet{Sober:1998:a}, which causes the appearance of group selection to vanish.  For example, measuring the proportion of cooperators in the vicinity of each and every cell or within its radius of influence will bias measurements of local proportions in such a manner that dense areas contribute more to the average in exact proportion to how dense they are -- in this case, the average local proportion cannot be different from the global proportion.

In the remainder of this paper we develop a simple individual-based model of bacterial growth, such as would apply to a locally-dispersing `public good', to illustrate the use of this methodology and as a basis for discussion of several additional complicating factors that are important in its application. Of particular interest is the possibility of measuring the local proportions at several different spatial scales to determine the effective scale of selection. 

\section{An individual-based model}
\subsection{Bacterial Biofilms}
In developing the following model we have bacterial biofilms in mind. Social evolution in bacterial systems is currently receiving considerable attention both as a model system of social evolution and because of the practical implications of biofilms \citep{Crespi:2001:a,Griffin:2004:a,Burmolle:2006:a}. Biofilms show a physical structure especially suited for localised fitness interactions via the formation of semi-isolated micro-colony structures \citep{Hall-Stoodley:2004:a}. However, the following model is general -- not dependent on any of the particulars that pertain to specific bacterial strains or types of fitness interaction. The vital assumptions are that there are two types of individual, that the presence of one of these types (but not the other) is beneficial to other individuals within a certain spatial radius, and that this type bears a cost for providing this benefit. For example, one type may be a wild-type strain of \textit{Pseudomonas Aurigenosa}, that releases into the environment an enzyme useful for binding iron \citep{Griffin:2004:a}. This enzyme can be understood as a `public good' because it can be used by others within the diffusion radius of the molecule. The other type may be a selfish mutant strain that does not produce the public good and is therefore not burdened by its production, but can, like any other individual, benefit from the public good produced by cooperators. 

\subsection{Model definition}
\label{secModelDef}
The state of the model at any point in time is defined by a population of individuals each of which has a type (cooperate/selfish), an age, a location in continuous 2D space and a `reproductive potential'. Reproductive potential can be thought of as the resources the individual has accumulated over time. There is no explicit modelling of the public good, diffusion constants, extra-cellular matrix, or such like - and in the default model, cells do not move. At every point in time, the fitness potential of each cell is incremented by a fitness benefit, $W$. This is a function of both the individual's own type, and of the number of cooperators in the local vicinity. Specifically, the fitness benefit of an individual is:

\begin{equation}
W=m+Pb-c,
\label{eqnFitness}
\end{equation}

where $m=1.5$ is a constant representing the intrinsic growth rate, $P$ is the proportion of cooperators within a given radius, $r_1=15$, of the individual (including itself), $b=4$ is a constant representing the fitness benefit received from cooperators, and $c=0$ for selfish individuals and $c=1.8$ for cooperators is the cost of being a cooperator (i.e., the cost of producing the public good). This fitness function is standard in evolutionary models of altruism \citep{Wilson:1980:a}, and amounts to an $n$-player public goods game / Prisoner's Dilemma \citep{Fletcher:2007:a}.

The model proceeds by updating each individual, in each time step, according to Algorithm 1.

\begin{algorithm}[h]
\begin{enumerate}
\item The age is incremented by 1.
\item	If the age is 5 the cell dies.
\item Otherwise, the fitness benefit is calculated (as above) and added to the individual's current reproductive potential.
\item	Whilst the reproductive potential $>4$,
\begin{enumerate} 
	\item	Reproduce, placing descendant cell in a new location according to a placement algorithm (see text). An offspring is an exact genetic clone of its parent.
	\item	Decrement reproductive potential by 4.
	\end{enumerate}
\end{enumerate}
\label{algUpdate}
\caption{Individual update algorithm.}
\end{algorithm}

The model is initialised with equal numbers of cooperators and selfish individuals distributed uniformly at random. Each initial cell (and new cell from reproduction) is initialised with reproductive potential=0, and age=0.
The placement algorithm may take account of competition for space (and possibly fail to produce an offspring if space does not allow) but by default it simply places an individual in a random location within a radius, $r_2=5$. Thus, an offspring is placed close to its parent.

Measuring the global proportion of cooperators is trivial. To measure the average local proportion of cooperators, the space is divided into contiguous square regions of size, $r_3=15$ (note that the area of each square local region, $(r_3)^2 =225$, in which local proportions are measured, is the same order of magnitude as the circular area over which a cooperator may affect other individuals, $\pi (r_1)^2 =707$. See Fig. 5.).

In an advanced version of the model, cells are motile and move toward cooperators. This represents attraction towards concentration of the public good, for example. At each time step, a vector is calculated which is a distance-discounted sum of vectors to all other local regions, weighted by the number of cooperators in that region. The regions used are the same as those used for calculating the average local proportion of types. Each cell then moves a random distance $d$ in the direction of this vector; $d$ is uniformly distributed in the range 0 to $15r_4$, where $r_4$ is a constant controlling the amount of movement.  

\section{Model illustrations}
  
We initialised each simulation with 150 cooperators and 150 selfish individuals, distributed randomly across a square grid of size $250*250$. Each simulation was repeated 50 times, and the mean of both the average local and global proportions of cooperators recorded. 

\begin{figure}[h]
\centering
\includegraphics[scale=0.35]{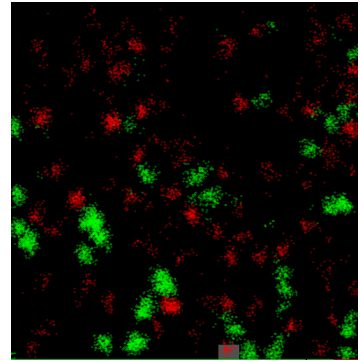}
\caption{Illustration of biofilm growth in the model. Green cells are cooperators, red are selfish cheats.}
\label{figScreenshot}
\end{figure}

Figure~\ref{figScreenshot} shows that although the initial distribution of bacterial cells is random, the cells grow into spatial clusters due to non-motility and the fact that offspring are placed close to their parents (as per \textit{Model definition}). 

\begin{figure}[!hb]
\centering
\subfloat[]{\label{figNoParadox}\includegraphics[scale=0.25]{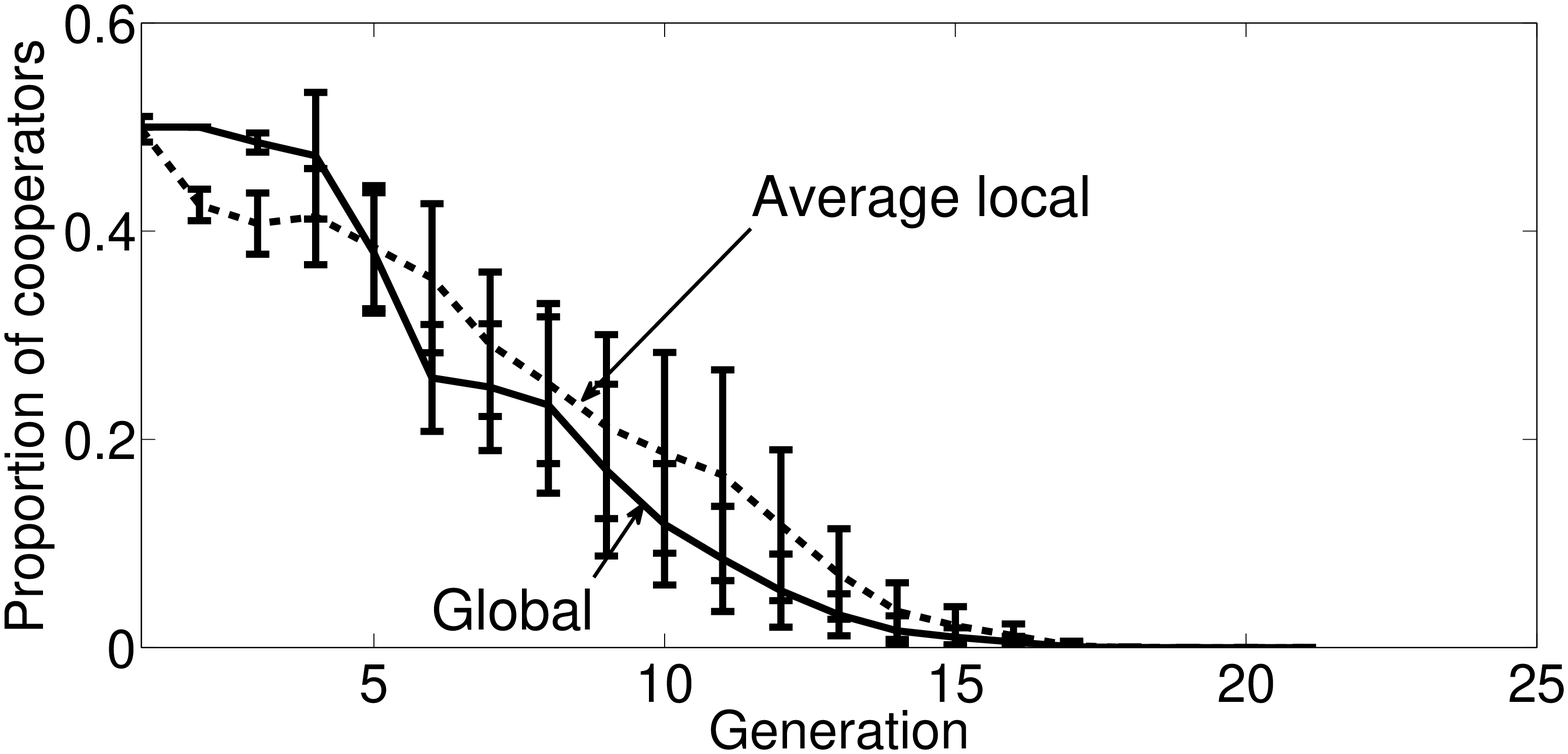}}
\\ \subfloat[]{\label{figParadox}\includegraphics[scale=0.25]{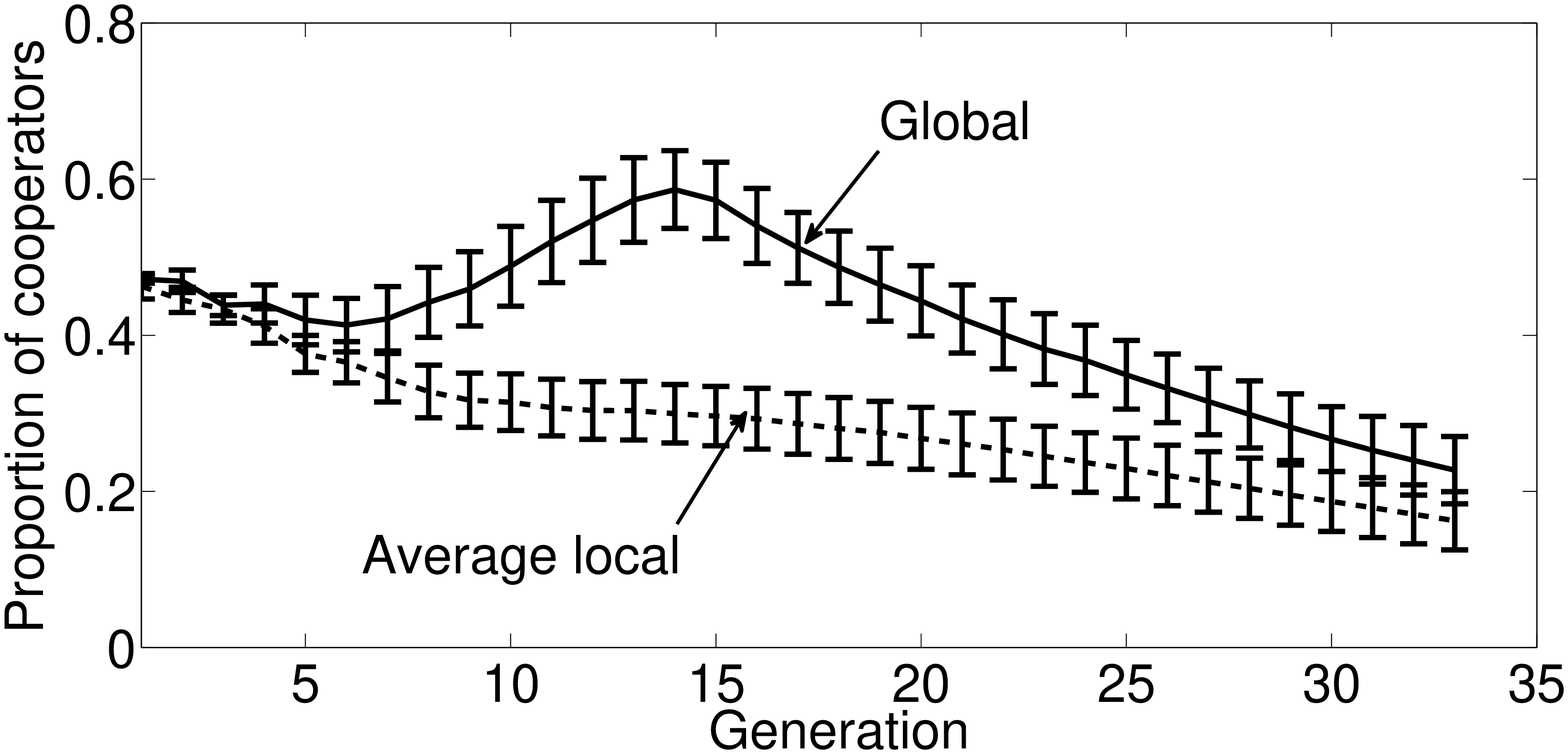}}
\\ \subfloat[]{\label{figAggDisp}\includegraphics[scale=0.25]{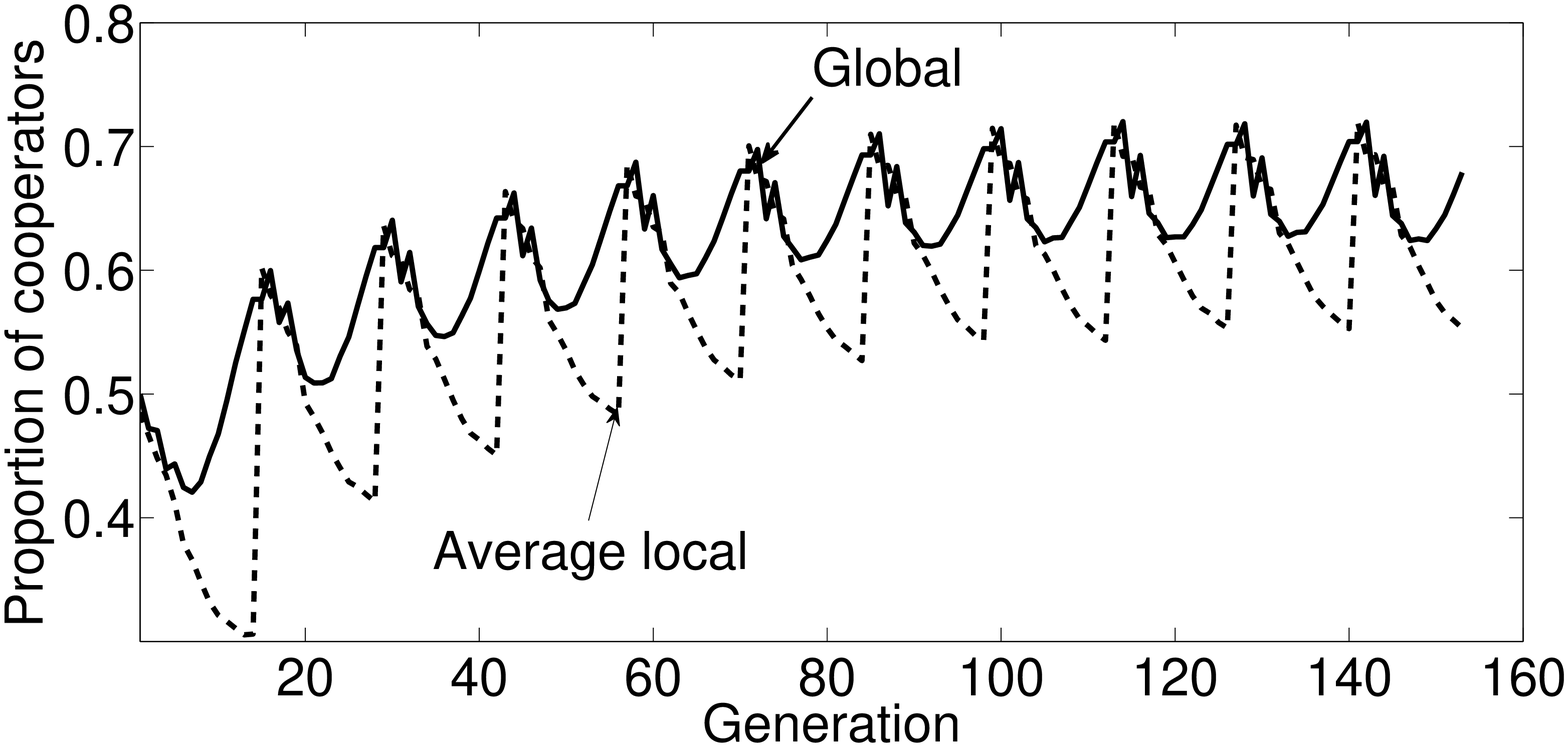}}
\caption[Local and global proportions of cooperators over time. Error bars show standard deviation.]{A) When the average of interaction, $r_1$ covers the entire space, cooperation does not evolve and Simpson's Paradox is not observed. B) When $r_1=15$ cooperation evolves, and there is a difference between local and global proportions. C) Multiple aggregation and dispersal cycles with $r=15$.}
\label{figNewPlot}
\end{figure}

From standard social evolution theory, we would not expect cooperation to increase or be stable in the absence of localised interactions \citep{Wilson:1980:a}. Thus, in such cases we should not see a Simpson's Paradox, since without localised interactions there should be no difference in the growth-rates of different localities, \textit{ceteris paribus}. We verified that this was the case in our model by making the radius of social interactions, $r_1$, equal to the size of the whole grid. Thus, each individual would experience the global proportion of cooperation for the purposes of determining their fitness. This corresponds to complete mixing of the public good, but not of the individuals themselves. Thus, we still measured the local proportion of cooperation across squares of size $r_3=15$. As Figure~\ref{figNoParadox} shows, the global frequency of cooperation steadily declines in this case, and there is no observation of a Simpson's Paradox. This is because although there are still spatial groups in the system, membership of these groups does not affect fitness when the public good is global, and hence they are meaningless to evolution. This serves as an illustration of the fact that the groups we can readily observe in a system (e.g., the clusters in our model) may not be the same scale as the groups that matter for the evolution of cooperation (in the case of well-mixed public goods, the `group' is the whole population).

On the other hand, in Figure~\ref{figParadox} we set the radius of the public good to $r_1=15$. This represents localised interactions, and so we might expect cooperation to evolve. Moreover, we set the window size over which we measure local proportions of cooperators to be of this same scale ($r_3=15$). In this case cooperation evolves, and we observe a difference between average local and global proportions of cooperation, and hence a Simpson's Paradox. It should be noted that Simpson's Paradox is present even when the global proportion of cooperators is falling, so long as the average local proportion of cooperators is falling at a faster rate (e.g., generations 1-6 in Figure~\ref{figParadox}). In this case there is a non-zero between-group component of selection, but this is weaker than within-group selection.

Figure~\ref{figParadox} also illustrates that the paradox cannot be sustained indefinitely. This is because selfish individuals are fitter than cooperators sharing the same public good (same $P$ value but $c=0$ in Equation~\ref{eqnFitness}). Thus, they must necessarily increase in frequency within each locality. As this happens, the differential growth of different localities decreases, and hence the paradox reduces. In Figure~\ref{figParadox}, the paradox peaks at 14 generations, after which the global frequency of cooperation starts to fall back down. This seemingly inevitable decrease in cooperation as the generations go by need not occur, however, if individuals are periodically mixed and redistributed in space \citep{Sober:1998:a}. Essentially this is because such a redistribution of individuals reestablishes variance in the proportion of cooperators (and hence in the amount of the public good) between groups, and so once again allows for differential group productivity to have an effect and create a paradox.  This is illustrated in Figure~\ref{figAggDisp}, where dispersal from clusters and global mixing occurs every 14 generations. These dispersal events explain the see-saw shape of the average local curve: at each dispersal event, the average local proportion is returned to the global proportion of cooperators. Dispersal is known to occur in natural biofilms \citep{Ghannoum:2004:a} (although simultaneous and complete mixing is a simplifying assumption of our model), and the single-celled bottleneck in the development of multicellular organisms provides a similar redistribution of genetic variance \citep{Smith:1995:a,Michod:1999:a}. Thus, some degree of dispersal is likely to be important in maintaining cooperation in natural populations \citep{West:2002:a}, and may actually be an evolutionary adaptation at least partly for this purpose \citep{Smith:1995:a,Michod:1999:a}.

\begin{figure}[h]
\centering
\includegraphics[scale=0.25]{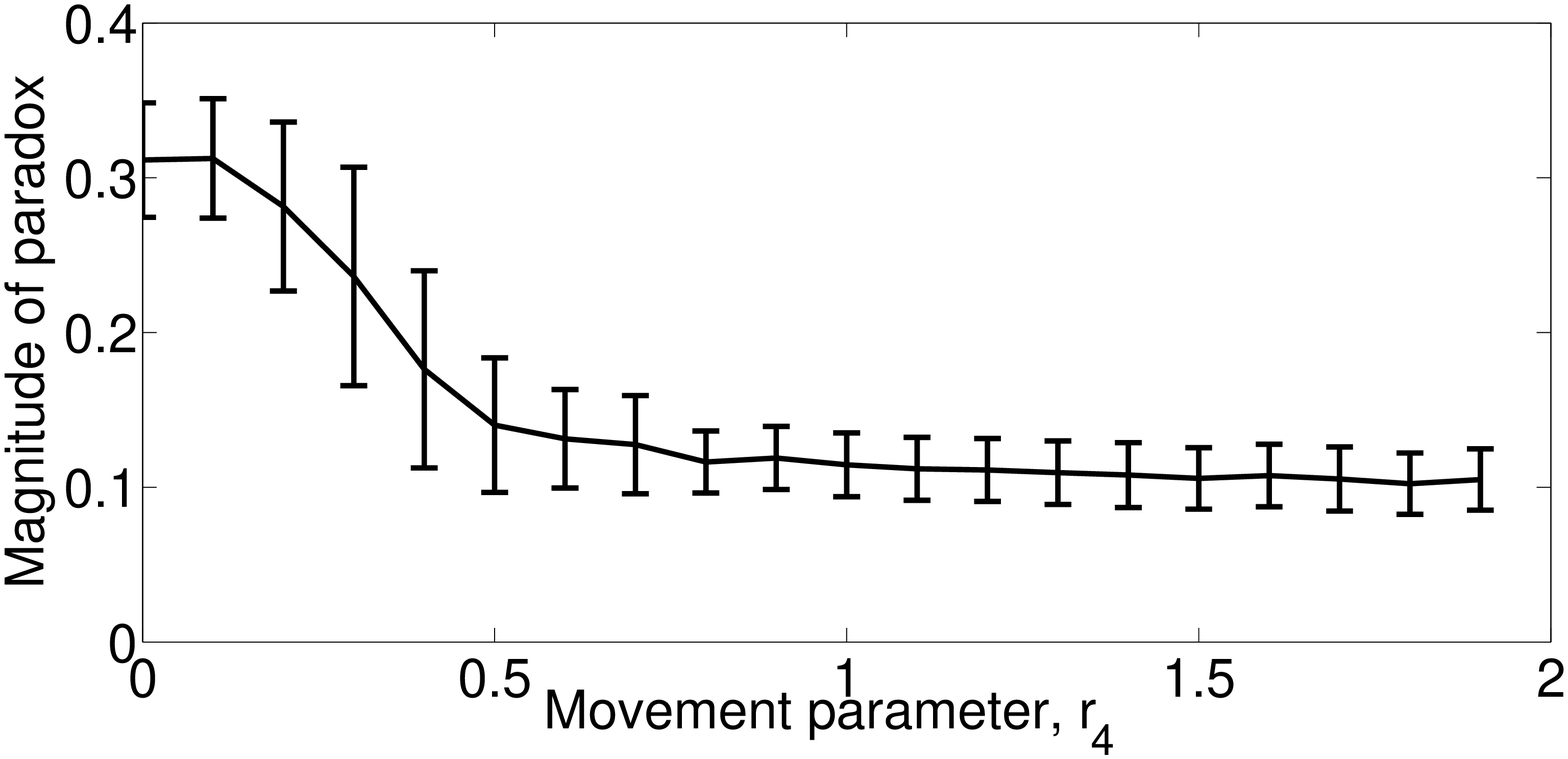}
\caption{Effect of increasing cell motility on the observation of Simpson's Paradox. Error bars show standard deviation.}
\label{figMovement}
\end{figure}

Figure~\ref{figMovement} shows the effect of cell motility on the observation of Simpson's Paradox. Again, from standard theory we would expect increasing motility to reduce global levels of cooperation. We see that increasing motility decreases Simpson's Paradox. This is because it increases the heterogeneity of localities, making their $P$ values more similar and hence the differential in group productivity lower. 

\begin{figure}[h]
\centering
\includegraphics[scale=0.25]{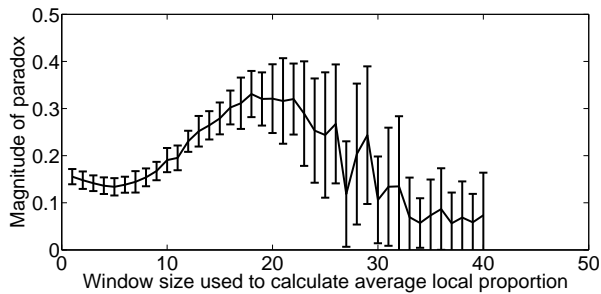}
\caption{Effect of the magnitude of the locality size measured on the observation of Simpson's Paradox (difference between local average and global proportion of cooperators). The observed paradox is strongest when the measured locality size corresponds to the actual scale of social interaction; measurements were taken after the number of generations that yielded the peak difference between global and local frequencies, for each window size. Error bars show standard deviation. The length of the error bars increases with the window size because a larger window size corresponds to fewer localities and hence fewer samples to average over.}
\label{figLocalitySize}
\end{figure}

Figure~\ref{figLocalitySize} shows how the peak observation of a Simpson's Paradox changes depending on the scale at which local proportions of cooperators are measured. Observation of the paradox will peak when this scale corresponds to the actual scale of social interactions in the system, e.g., to the radius in which the public good is shared. The peak in Figure~\ref{figLocalitySize} is where the measured locality size corresponds, approximately, to $r_1$, the actual scale of interaction. Measuring Simpson's Paradox using different local scales could thus be used to determine the actual scale of social interactions in a real-world system, where this may well not be known \textit{a priori}.

\section{Discussion}

We have presented a methodology for measuring the effect of group-level selection in natural populations. Real-world populations may often not be formed of clearly observable groups with discrete boundaries, which makes the application of standard multilevel selection theory non-trivial. In particular, theoretical techniques for measuring the strength of group selection, such as the Price Equation or contextual analysis, rely on being able to measure properties of discrete groups \citep{Godfrey-Smith:2006:a}. Thus, their application to systems such as bacterial biofilms remains problematic.

Here, we have suggested observation of Simpson's Paradox as a way to quantify the effect of group-level selection in a natural population. It is now widely appreciated that Simpson's Paradox, the difference between average local and global frequencies of cooperation, will be present whenever individually-costly cooperative behaviours evolve \citep{Sober:1998:a}. Moreover, its presence indicates multiple scales of selection in a system \citep{Sober:1998:a}. However, discussions of Simpson's Paradox have so far remained in the theoretical domain. In particular, illustrations of it have, to our knowledge, only been conducted in models with discrete group boundaries. By contrast, we have shown that Simpson's Paradox can be readily measured in populations where individuals are continuously distributed throughout space. Thus, the exact group structure does not have to be known \textit{a priori} for this technique to be applied. We have illustrated the measurement of Simpson's Paradox in such a case with an individual-based model of public goods production in bacterial biofilms.

Significantly, measurement of Simpson's Paradox can be used to determine the effective group structure in a natural population. Specifically, the difference between average local and global proportions of cooperation will peak when the size of localities measured is of the same scale as that over which the public good is shared. That is, when the measurement window size matches the scale of fitness-affecting social interactions. \citet{Wilson:1980:a} terms the scale over which social interactions occur ``trait groups''. He stresses that the groups which matter to natural selection are subsets of individuals in which fitness-affecting interactions occur, and that these subsets may not correspond to the apparent groups that are most readily observable in a population. For example, although discrete clusters may be observable in a biofilm, these may not correspond to the radius over which a public good diffuses. Varying the window size over which the change in local proportions of cooperators is measured, and looking for the peak difference with the global proportion, can identify the effective trait groups in the population. Searching for the trait groups in this way can be done by image analysis at the end of the experiment -- the experiment does not have to be re-run in order to measure Simpson's Paradox on different scales. Regarding biofilms, one may also measure local proportions using regions that specifically enclose micro-colonies to see if micro-colony structure is a stronger selective unit than arbitrary local regions. That is, our methodology can be used to determine whether the micro-colonies correspond to trait groups, or whether the trait groups are in fact smaller or larger. 

In future work, it would be interesting to investigate whether the Price Equation can be meaningfully applied to the appropriate window size. In particular, our methodology identifies non-arbitrary groups. Thus, once we have identified the effective trait group size, we could calculate the covariance between group character (local proportion of cooperators), and group productivity. Likewise, we could calculate the covariance between individual character (cooperator or not) and individual fitness (number of cell divisions). Our methodology also fits within a kin selection framework \citep{Hamilton:1964:a}, as used by \citet{Griffin:2004:a} to study bacterial social evolution, for example. Finding the trait groups corresponds to finding the scale at which genetic relatedness should be measured in a natural population.

\subsubsection*{Acknowledgements}
Thanks to Alex Penn, Jeremy Webb and Lex Kraaijeveld.


\end{document}